\documentclass[useAMS,usenatbib]{mnras}

\usepackage{color}

\usepackage{times}

\usepackage{graphicx}	
\usepackage{amsmath}	
\usepackage{amssymb}	

\newcommand{\be}{\begin{equation}}
\newcommand{\ee}{\end{equation}}
\newcommand{\beq}{\begin{eqnarray}}
\newcommand{\eeq}{\end{eqnarray}}

\newcommand{\fc}{f_{\rm c}}

\title[Helium accreting neutron star in 4U\,1820$-$30]{Basic parameters of the helium accreting X-ray bursting neutron star in 4U\,1820$-$30}

\author[V. F. Suleimanov et al.]{Valery F. Suleimanov,$^{1,2,3}$\thanks{E-mail: suleimanov@astro.uni-tuebingen.de}
Jari J.~E. Kajava,$^{4,5}$
Sergey V. Molkov,$^3$
Joonas N\"attil\"a,$^{5,6}$
\newauthor
Alexander A. Lutovinov,$^3$
Klaus Werner,$^{1}$
and 
Juri Poutanen$^{5,6}$
\\
$^{1}$Institut f\"{u}r Astronomie und Astrophysik, Kepler Center for Astro and
Particle Physics, Universit\"{a}t T\"{u}bingen, Sand 1, D-72076 T\"{u}bingen,
Germany\\
$^2$Kazan (Volga region) Federal University,  Kremlevskaya str. 18, Kazan 420008, Russia\\
$^3$Space Research Institute, Russian Academy of Sciences, Profsoyuznaya 84/32,
117997 Moscow, Russia\\
$^{4}$Finnish Centre for Astronomy with ESO (FINCA), University of Turku, V\"{a}is\"{a}l\"{a}ntie 20, FIN-21500 Piikki\"{o}, Finland\\
$^{5}$Tuorla Observatory, Department of Physics and Astronomy, University of Turku, V\"{a}is\"{a}l\"{a}ntie 20, 
FIN-21500 Piikki\"{o}, Finland\\
$^{6}$Nordita, KTH Royal Institute of Technology and Stockholm University, Roslagstullsbacken 23, SE-10691 Stockholm, Sweden}

\pubyear{2016}

\begin{document}
\label{firstpage}
\pagerange{\pageref{firstpage}--\pageref{lastpage}}
\maketitle

\begin{abstract}
The ultracompact low-mass X-ray binary 4U\,1820$-$30 situated in the globular cluster NGC\,6624 has 
an orbital period of only $\approx 11.4$\,min which likely implies a white dwarf companion.   
The observed X-ray bursts demonstrate a photospheric radius expansion phase and therefore  
are believed to reach the  Eddington luminosity allowing  us to estimate the mass and 
the radius of the neutron star (NS) in this binary.  
Here we  re-analyse all {\it Rossi X-ray Timing Explorer}  observations of the system 
 and confirm that almost all the bursts took place during the hard persistent state of the system. 
This allows us to  use the recently developed direct cooling tail method to estimate the NS mass and radius.  
However, because of the very short, about a second, duration of the cooling tail phases 
that can be described by the theoretical atmosphere models, the  obtained constraints on the NS radius  are not very strict.  
Assuming a pure helium NS atmosphere we found that
the NS radius is in the range 10--12\,km, if the NS mass is below 1.7\,M$_\odot$, and in 
a wider range of 8--12\,km for a higher 1.7--2.0\,M$_\odot$ NS mass. 
The method also constrains the distance to the system to be 6.5$\pm$0.5\,kpc, which is consistent with the distance to the cluster.
For the solar composition atmosphere,  the NS parameters are in strong contradiction 
with the generally accepted range of possible NS masses and radii.  
\end{abstract}



\section{Introduction}

Many accreting neutron stars (NSs) in low-mass X-ray binaries (LMXBs) are sources of relatively bright X-ray bursts, 
extensively investigated by the {\it Rossi X-ray Timing Explorer (RXTE)}  in the past two decades \citep[see, e.g.,][]{gallow08}. 
The X-ray bursts are produced by thermonuclear flashes on NS surfaces of the freshly accreted matter from 
the binary companion \citep[see, e.g., reviews by][]{LvPT93, bildsten98, cumming04, SB06}.
The emergent spectra of X-ray bursting NSs are usually well fitted with a blackbody \citep{gallow08} because of 
the importance of the Compton effect in their atmospheres
 \citep{Londonetal:84, Londonetal:86, Lapidusetal:86, Ebisuzaki87,Pavlov.etal:91, madej:91}. 
Some of the powerful bursts reach the Eddington luminosity and this fact is very useful for limiting  the NSs masses 
$M$ and radii $R$ \citep{Damen:90, LvPT93, Ozel06}.
These bursts demonstrate an appearance of the radius expansion phase \citep{PA86, Damen:90, LvPT93}, 
and they are called photospheric radius expansion (PRE) bursts.
 
Masses and radii of NSs comprise important information about the physics of cold superdense matter in their cores
 \citep[see, e.g.,][]{HPY:07, LP07, LS:14}. 
Properties of such matter are not well understood yet and this issue is known as the problem of the equation of state (EoS). 
A few years ago, a novel cooling tail method for determination of masses and radii of X-ray bursting NSs was
 suggested \citep{SPRW11}.
This method is based on the approaches developed before by \citet{Ebisuzaki87} and  \citet{vP:90}, and on  modern
model atmospheres of X-ray bursting NSs \citep{SPW11, SPW12}. 
It was recently modified to remove some  systematic uncertainties present in the original method \citep{Sul.etal:17}. 
The method assumes that a burst can be described by a passively cooling uniform NS without any influence of the accreted matter.
Such conditions seem to be fulfilled only in PRE bursts taking place during a low/hard persistent state 
of the host LMXB  \citep{SPRW11, Kajava.etal:14, Suleimanov16EPJA}. 
The method was applied to some X-ray bursts \citep{SPRW11, Poutanen.etal:14, nattila16}. 
We note, however, that the method can be used for non-PRE bursts during hard persistent state as well \citep{Zamfir:12}.

The main uncertainty of the method is the chemical composition of the X-ray bursting NS atmosphere.
It is mainly determined by the chemical composition of the accreted matter, which is actually unknown, and the results
 obtained using hydrogen-rich and hydrogen-poor model atmospheres differ significantly \citep{SPRW11, Poutanen.etal:14}.
There are two reasons for that. 
Firstly, the model curves $w - w\fc^4\,L/L_{\rm Edd}$, which have to be compared with the observed curves 
$K - F_{\rm BB}$, depend on the chemical composition. 
Here $L$ is the luminosity, $L_{\rm Edd}$ is the  
Eddington luminosity of the considered NS, $K$ is the normalization factor of a blackbody fitted to the observed spectrum
 $F_{\rm E}$,  $F_{\rm BB}$ is the blackbody observed flux, and $E$ is the photon energy. 
The dilution factor $w$ and the color correction factor $\fc$ are the parameters of the  diluted blackbody spectrum 
$\mathcal{F}_{\rm E} \approx w\,\upi B_{\rm E}(\fc T_{\rm eff})$ fitted to the data.  
The colour temperature $T_{\rm c}=\fc\,T_{\rm eff}$ is larger than the effective temperature $T_{\rm eff}$ by 
the colour-correction factor $\fc > 1$.
Secondly, the Eddington luminosity depends on the hydrogen mass fraction $X$, 
$L_{\rm Edd} = 4\upi\,GM\,c\,(1+z)/0.2\,(1+X)$, where $z$ is the surface gravitational redshift.  
As a result, the solution obtained assuming pure helium atmosphere gives twice as large NS radii in comparison with
the solution for a pure hydrogen atmosphere.

The photospheric chemical composition of X-ray bursting NSs can be potentially enriched by heavy elements 
because of thermonuclear burning at the bottom of the accreted envelope  \citep{Weinberg.etal:06, intZW10}.
To investigate the effect of metal enrichment on the emergent atmosphere spectra, an extensive set of model
 atmospheres with various heavy element abundances (up to pure iron atmospheres) were recently computed 
 by \citet{Netal15}, see also \citet{Rauchetal:08}. 
The main conclusion of this work is that the colour correction $\fc$ decreased with increasing heavy element 
abundances at a given relative model luminosity $L/L_{\rm Edd}$. 
Importantly, a significant jump in the $K - F_{\rm BB}$ curve has been observed in  the X-ray burster HETE\,J1900.1$-$2455, 
which was interpreted as a transition from the metal-enriched photosphere to a photosphere covered by the freshly accreted 
matter with a solar H/He mix \citep{Kajava17}.
Moreover, an absorption edge was also found in the spectra of HETE\,J1900.1$-$2455 at the suggested metal-enriched
 X-ray burst phase.  
Unfortunately, relatively little heavy element enhancement  (up to ten times solar abundance) cannot be easily detected 
and it may lead to a large source of systematic uncertainty for  NS mass and radius measurements using X-ray bursts. 
For example, this was the likely reason for the overestimation of the NS radius  in 4U\,1724$-$307 by \citet{SPRW11}, 
when compared with the results obtained later by \citet{nattila16} for a different burst of the same source.

The significant difference  between obtained NS radii, when hydrogen-rich  and helium-rich model atmospheres are used, 
offers the possibility  to distinguish between NSs accreting matter of corresponding chemical composition. 
In particular, an acceptable solution for the X-ray bursting NS in 4U\,1702$-$429 was only obtained using a pure helium 
atmosphere  \citep{nattila16}. 
The orbital period of that system is unknown and, therefore, there is no direct evidence that the system is ultracompact 
binary like 4U\,1820$-$30 that is known to have a helium-rich white dwarf companion. 
However, there is indirect evidence such as a relatively large ratio of the energy release between bursts to the burst
 fluence ($\alpha \approx 75$), and  relatively short burst rise times and flux decay time scales \citep{gallow08}. 
It is very important to apply the direct cooling tail method to the X-ray bursts of 4U\,1820$-$30 to validate the 
method using a system with a helium-rich companion.
This is the main aim of the present work. 
 
The ultracompact X-ray binary 4U\,1820$-$30 with an orbital period of 685~s \citep{SPW87} is located in 
the globular cluster NGC\,6624 at a distance of 7.6$\pm$0.4\,kpc \citep{Heasley.etal:00, Kuulkers.etal:03}.
Most probably, the LMXB 4U\,1820$-$30 is a member of a triple system, as the persistent X-ray flux of the system
 is modulated  with a period $\approx$\,171\,days  \citep{PT84, CG01,ZWG07}. 
Thermonuclear X-ray bursts in the system discovered by the {\it ANS} observatory \citep{grindlay76}
 occur near the minimum of  the persistent X-ray flux \citep{Clark.etal:77, CG01}. 
It is commonly accepted that the secondary star in the system is a helium white dwarf with a mass of
 $\approx$\,0.06--0.07 M$_\odot$ \citep{Rappoport.etal:87}. 
The hypothesis of an almost pure helium composition of the accreted matter is suggested by a large value 
of  $\alpha \approx 125 - 155$ obtained  by \citet{Haberl.etal:87} using {\it EXOSAT} observations. 
A detailed comparison of theoretical predictions derived from  thermonuclear flash models with observations 
was performed by \citet{C03}, who showed that the hydrogen mass fraction in the accreted matter cannot exceed 0.1.  
One long superburst connected with carbon burning was also detected in the system \citep{SB02}.
 
Thermonuclear X-ray bursts of 4U\,1820$-$30 were intensively investigated using the {\it RXTE} observatory 
\citep[see, e.g.][]{Guver.etal:10},  and the most detailed investigation was made by \citet{GZM13} \citep[see also][]{Kajava.etal:14}. 
They found 16 short PRE X-ray bursts with a total duration of about 20--30~s
 and showed that all of them (with one exception) took place in a hard spectral state. 
Moreover, they  demonstrated, that the model curve $\fc - L/L_{\rm Edd}$ computed for pure helium atmospheres describes 
reasonably well the observed curve $K^{-1/4}-F_{\rm BB}$ in the luminosity interval   $L/L_{\rm Edd}\sim 0.55 - 0.95$.  
Therefore, we can expect that the direct cooling tail method can be used for NS mass and radius determination in this system.
 
\section{Analysis of the observational data}
\subsection{Persistent emission}

\begin{figure} 
\begin{center}
\includegraphics[width= 0.9\columnwidth]{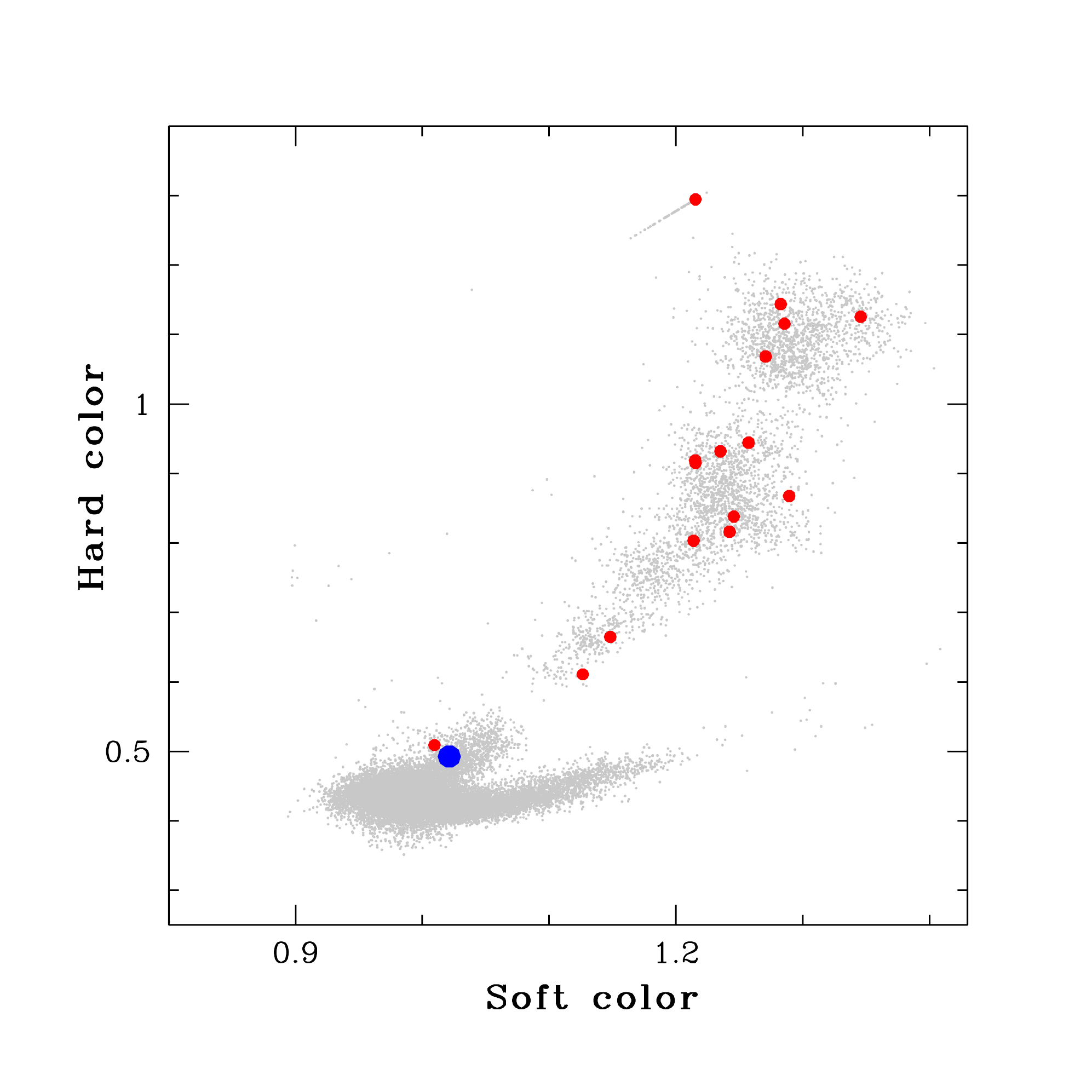}
\caption{\label{fig:color} 
Colour-colour  diagram for the persistent spectrum of 4U\,1820$-$30 as observed by {\it RXTE} in all separate 48~s time windows. 
The position of the source before short X-ray bursts are marked by the red dots.  
The superburst  is marked by the blue dot.
} 
\end{center} 
\end{figure}

We analysed all available {\it RXTE}/PCA observations of 4U\,1820$-$30 using standard reduction procedures, described
in detail in \citet{Kajava.etal:14}. 
There are 16 PCA observations where short PRE X-ray bursts are seen, and they have been studied earlier 
by \citet{GZM13} and \citet{Kajava.etal:14}. 
All observations were divided into 48~s long intervals, making together about 27\,000 observing windows. 
To investigate their spectral properties the observed energy band was divided into three bands, 3--6 keV, 6--12 keV and 12--25 keV, and the corresponding fluxes $F_{3-6}$, $F_{6-12}$ and $F_{12-25}$ were estimated using a two-component model described below. 
Using the soft (SC=$F_{3-6}/F_{6-12}$) and the hard (HC=$F_{12-25}/F_{6-12}$) colours we obtained a colour-colour diagram (Fig.\,\ref{fig:color}), which is similar to those obtained by \citet{GZM13} and \citet{Kajava.etal:14}. 
As it was shown in the cited works, the bursts occur  predominantly in the hard persistent states (the hard colour is larger than 0.5). 

\begin{figure} 
\begin{center}
\includegraphics[width= 0.9\columnwidth]{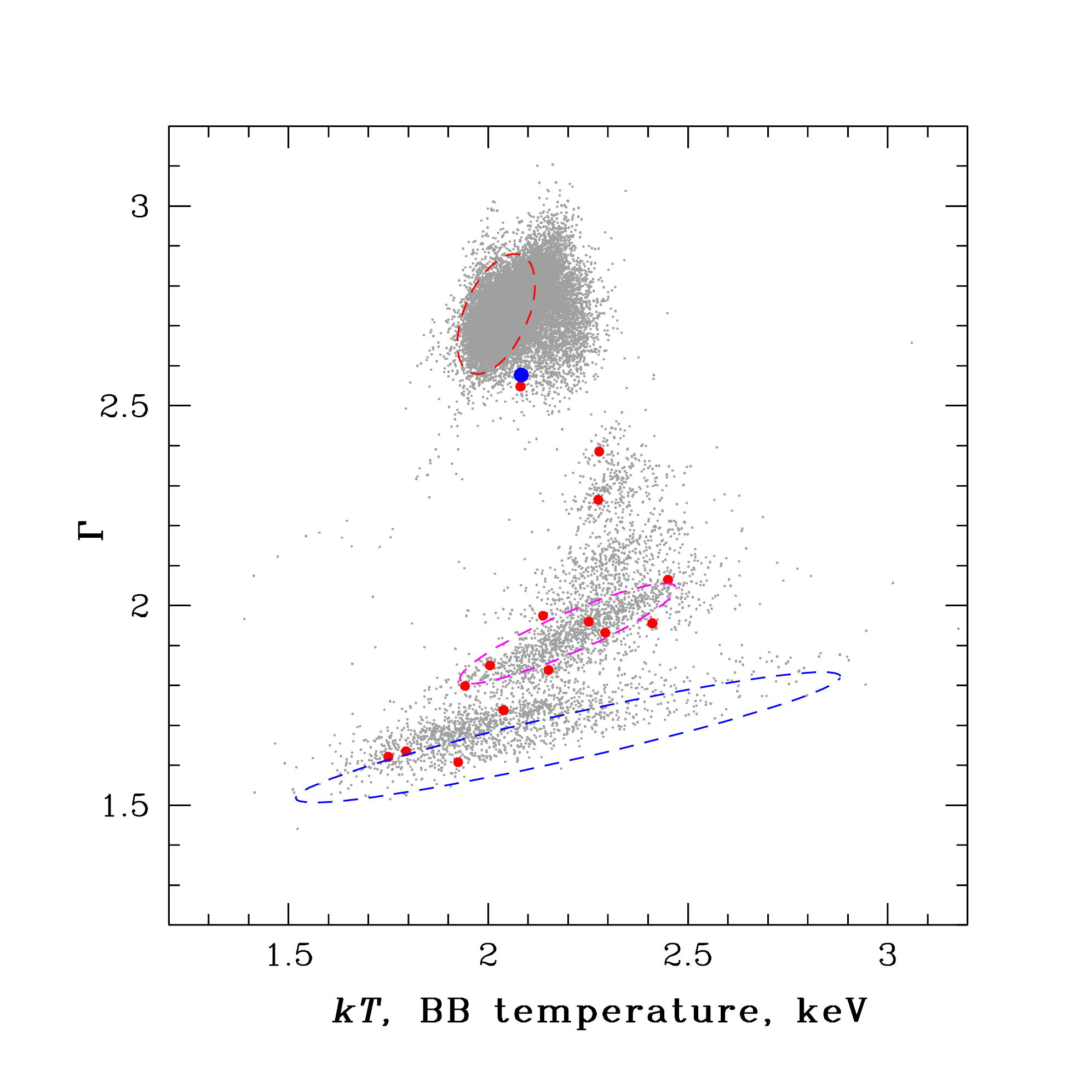}
\caption{\label{fig:gammakT} 
Same  as Fig.~\ref{fig:color} but for  the $\Gamma - kT$ dependence. 
 The spectra assigned to three
different groups are marked by different colour dashed ovals: the top red one represents the soft state, the middle magenta one
represents the intermediate  state and the bottom blue one represents the hard state. 
} 
\end{center} 
\end{figure}

The persistent spectra were analysed using the {\sc xspec} \citep{Arnaud96} package and were fitted with a two-component   {\sc bbodyrad + powerlaw} model with the main fitting parameters being the blackbody temperature $kT$ and the power-law index $\Gamma$. 
Additionally the interstellar absorption was accounted for with model {\sc wabs} with the fixed value of the hydrogen column density of $N_{\rm H} = 1.6 \times 10^{21}$\,cm$^{-2}$ following \citet{Costantini12}.   
 The $\Gamma - kT$ diagram is shown in Fig.\,\ref{fig:gammakT}.  
 We note that the power-law component dominates in the hard persistent states when the hard color $> 0.5$ and $\Gamma < 2.5$. 
 The contribution of the blackbody component decreases with a decreasing $\Gamma$. 
 As a result, the blackbody temperature becomes very uncertain at low $\Gamma$ and ranged from 1.5 to 3 keV. 
 On the other hand, the contribution of the power-law component is insignificant in the soft persistent state (the hard colour is less than 0.5) and the spectra are almost blackbodies with  temperatures of about 1.9--2.3 keV.

\begin{figure} 
\begin{center}
\includegraphics[width= 0.9\columnwidth]{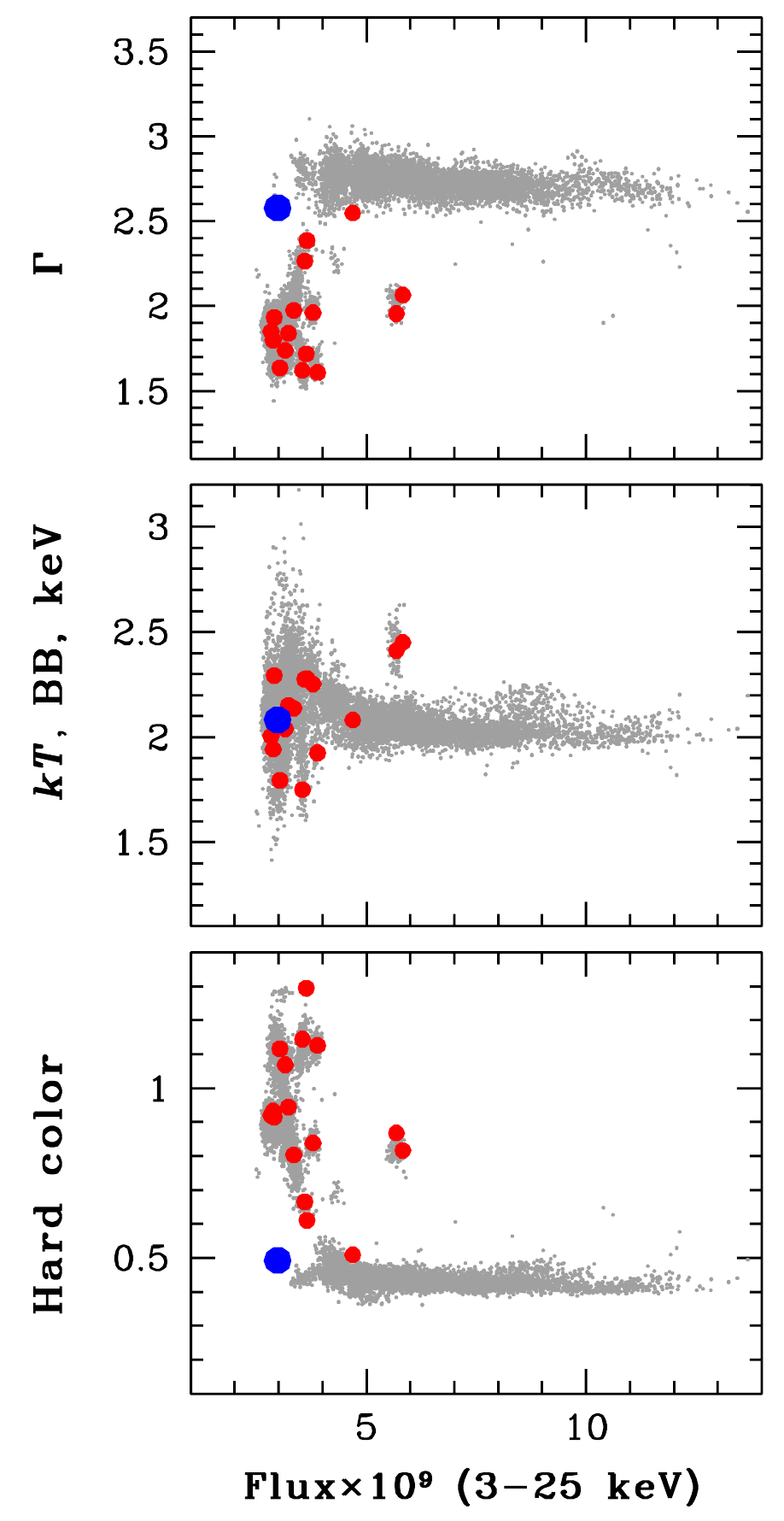}
\caption{\label{fig:all_flux} 
Dependences of $\Gamma$, $kT$, and the hard colour  on the  observed persistent flux (in units of $10^{-9}$~erg~cm$^{-2}$~s
$^{-1}$) in the 3--25 keV band  for all 48~s long observing windows of 4U\,1820$-$30. 
The positions of the source at the time of short X-ray bursts are marked by the red circles.  
} 
\end{center} 
\end{figure}

The dependences of three persistent state parameters, $kT$, $\Gamma$ and the hard colours on the observed flux in the 3--25 keV band are shown in Fig.~\ref{fig:all_flux}. 
This figure confirms a well known fact that the bursts in 4U\,1820$-$30 take place predominantly at low luminosities in the hard persistent state. 
However, one {\it RXTE} observation caught two  bursts from the source when it was at a significantly higher luminosity, but still  in the hard spectral state. 
This means that the possibility for the thermonuclear flash  depends on the persistent spectral state rather than just on the 
bolometric luminosity.    
 
\begin{figure}
\begin{center}
\includegraphics[width= 0.8\columnwidth]{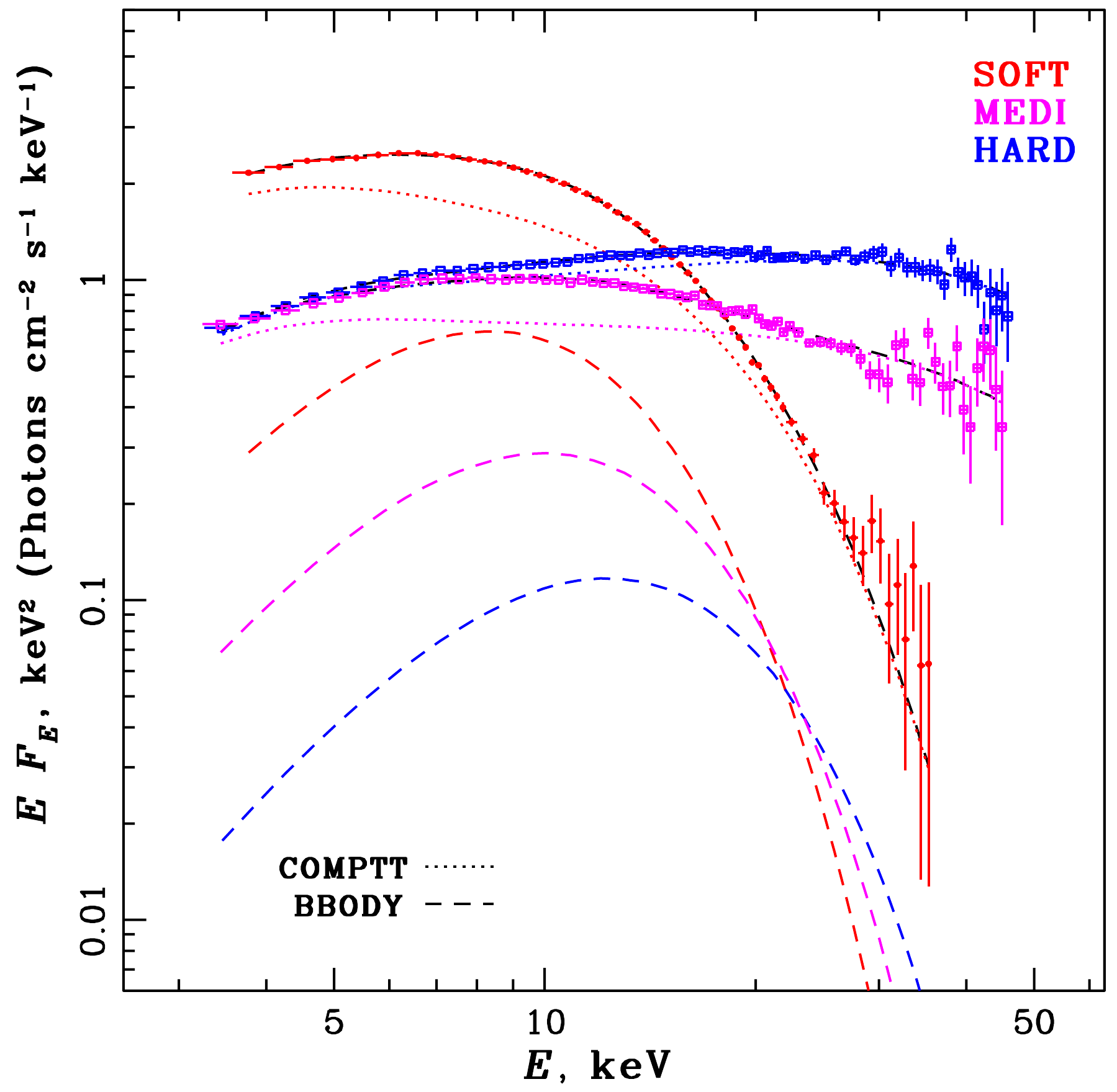}
\caption{\label{fig:aveper} 
The  persistent spectra of the source in three states (soft, intermediate and hard) averaged over some observations within the ovals marked in Fig.~\ref{fig:gammakT} together with the best-fitting two-component spectral models.  
} 
\end{center} 
\end{figure}

We performed a more detailed study of the persistent spectra. Some of them were summed into three groups shown in the $\Gamma - kT$ diagram (Fig.\,\ref{fig:gammakT}) by three different colours. 
They represent the soft, intermediate and hard spectral states. 
The averaged spectra of these three groups together with the best-fitting models  are shown in Fig.\,\ref{fig:aveper}. 
The spectra were fitted with the two-component  {\sc bbodyrad + comptt} model and same interstellar absorption as above. 
The obtained best-fitting parameters are presented in Table\,\ref{tab1}.  
We note, however, that this kind of spectral decomposition is ambiguous and the frequency-resolved spectroscopy has to be used to separate reliably different components \citep[see e.g. ][]{GRM03, RSP13}.  

 \begin{table}
\centering   
\caption{Best-fitting parameters for the averaged spectra of the three persistent spectral states. 
 \label{tab1} 
 }
\begin{tabular}[c]{l c c c }
\hline
Parameter & 	Soft        &  Intermediate    & Hard            \\ 
\hline 
$kT$           &  2.20$\pm$0.2   & 2.56$\pm$0.15   & 3.15$\pm$0.37   \\
$K$            &  8.3 $\pm$1.2   & 1.35$\pm$0.21   & 0.24$\pm$0.10   \\
$kT_0$         &  0.7$\pm$0.1    & 0.76$\pm$0.05   & 0.85$\pm$0.02   \\
$kT_{\rm e}$   &  3.6$\pm$0.4    & 12.35$\pm$3.1   & 12.0$\pm$1.3    \\
$\tau_{\rm e}$ &  4.2 $\pm$0.9   & 2.2 $\pm$0.5    & 2.77$\pm$0.25   \\
 \hline
\end{tabular}
\begin{flushleft}{ 
Note: $kT$ (keV) and $K$ are the temperature and normalization of the {\sc bbodyrad} model. 
$kT_0$ is the temperature (keV) of the seed photons, $kT_{\rm e}$ is the temperature (in keV) of the hot electrons, and $\tau_{\rm e}$ is the optical thickness of the electron slab for the {\sc comptt} model.
Errors are $1\sigma$.
}\end{flushleft} 
\end{table}

\subsection{Bursts}

\begin{figure} 
\begin{center}
\includegraphics[width= 1.\columnwidth]{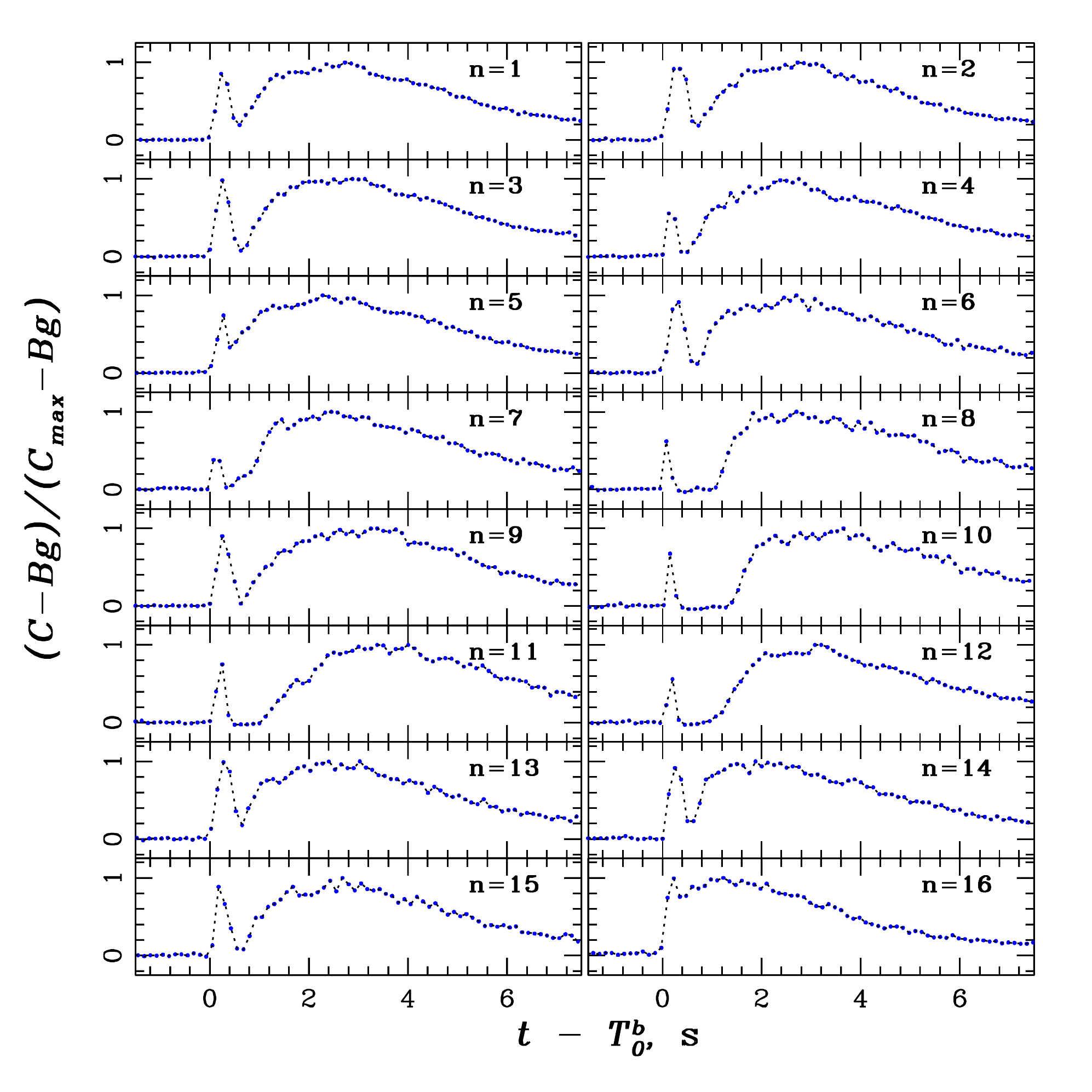}
\caption{\label{fig:lc} 
Light curves of the bursts in the energy band 3$-$25 keV normalized to the maximum  with the  persistent count rate subtracted as background.
} 
\end{center} 
\end{figure}

The light curves of all the short X-ray bursts of 4U\,1820$-$30 are shown in Fig.\,\ref{fig:lc}. 
All the bursts are very short with the bright phase lasting only about 10~s and all of them manifest clear PRE burst features with slightly different PRE phase durations.

The burst spectra were extracted as described in \citet{Kajava.etal:14}.
The burst light curves (Fig.\,\ref{fig:lc}) were divided into about 40 time bins with approximately the same number of photons 
in each of them.
The bin time duration therefore varies from 0.125~s at the brightest phases up to 4~s in the faintest time intervals.
The spectrum of each bin was fitted with the {\sc bbodyrad} model. 
The interstellar absorption taken into account with the {\sc wabs} model at fixed $N_{\rm H} = 1.6\times 10^{21}$\,cm$^{-2}$.
The burst spectra were also fitted with a slightly higher value of $N_{\rm H} = 2.5\times 10^{21}$\,cm$^{-2}$ as reported in \citet{Guver.etal:10}, but the results have not changed. 
The same conclusion was reached by \citet{GZM13}. 
In addition, in \citet{Kajava.etal:14} we fitted the bursts using $N_{\rm H}$ as a free parameter, and the results were found to be comparable.
Thus, the way in which we model the interstellar absorption does not influence the values of the blackbody model parameters.

What may influence the results is the choice of how we take into account corrections from the persistent spectrum. 
We used the standard approach, whereby the persistent spectra prior to the bursts were considered as 
a background and subtracted from the burst spectra. 
This method ignores the possible changes of the persistent spectra during the bursts
 \citep[see e.g.][]{Worpel.etal:13,Worpel.etal:15,Ji.etal:13}.
We therefore tested the presence of possible persistent spectral variations during the first burst
 (OBSID: 20075-01-05-00) with the method proposed by \citet{Worpel.etal:13}.
Here the persistent emission spectrum is not subtracted as a background, but instead its model spectrum
 with a free multiplier $f_\textrm{a}$ is added to the burst model spectrum and fitted to the data. 
In the cooling tail -- which we use to constrain the NS parameters -- 
the $f_\textrm{a}$ values were consistent with unity and the improvement to the fit was found to be insignificant.
As consequence the results did not change from the standard case, and therefore we decided to apply
 a standard procedure and subtract the persistent spectrum as in \citet{Kajava.etal:14}.

\begin{figure} 
\begin{center}
\includegraphics[width= 0.9\columnwidth]{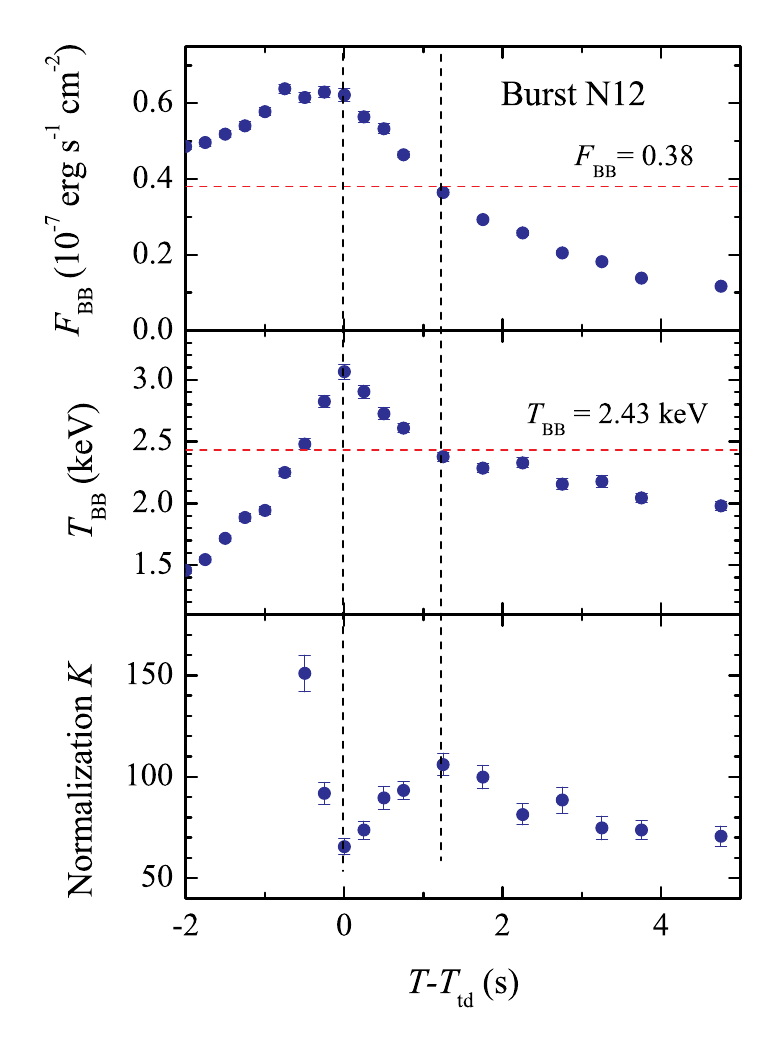}
\caption{\label{fig:cooling} 
Results of the blackbody fits to the burst \#12 (see also Fig.\,\ref{fig:lc} and Table\,\ref{tab2}).
The blackbody flux $F_{\rm BB}$, temperature $T_{\rm BB}$ and normalization $K$ in units (km\,/\,10\,kpc)$^2$ are shown.  
The points between the vertical dashed lines were used for the direct cooling method.
} 

\end{center} 
\end{figure}

An example of such a fit is shown in Fig.\,\ref{fig:cooling}. 
The touchdown point, where the blackbody temperature $T_{\rm BB}$ reaches the maximum and 
the normalization $K$ reaches the minimum, is marked by the left vertical dashed line. 
The normalization $K$ undergoes clear evolution after the touchdown point. It has a minimum value at the touchdown,
 then increases up to the maximum value after approximately one second, and then slowly decreases. 
The maximum value of the normalization approximately corresponds to $kT_{\rm BB} =2.4$\,keV and the blackbody 
 flux $F_{\rm BB} = 0.38\times 10^{-7}$\,erg s$^{-1}$ cm$^{-2}$. 
These values are close to the knee point in the $F_{\rm BB} - kT_{\rm BB}$ dependence, which was found by \citet{GZM13}.  
 
 \begin{figure} 
\begin{center}
\includegraphics[width= 0.9\columnwidth]{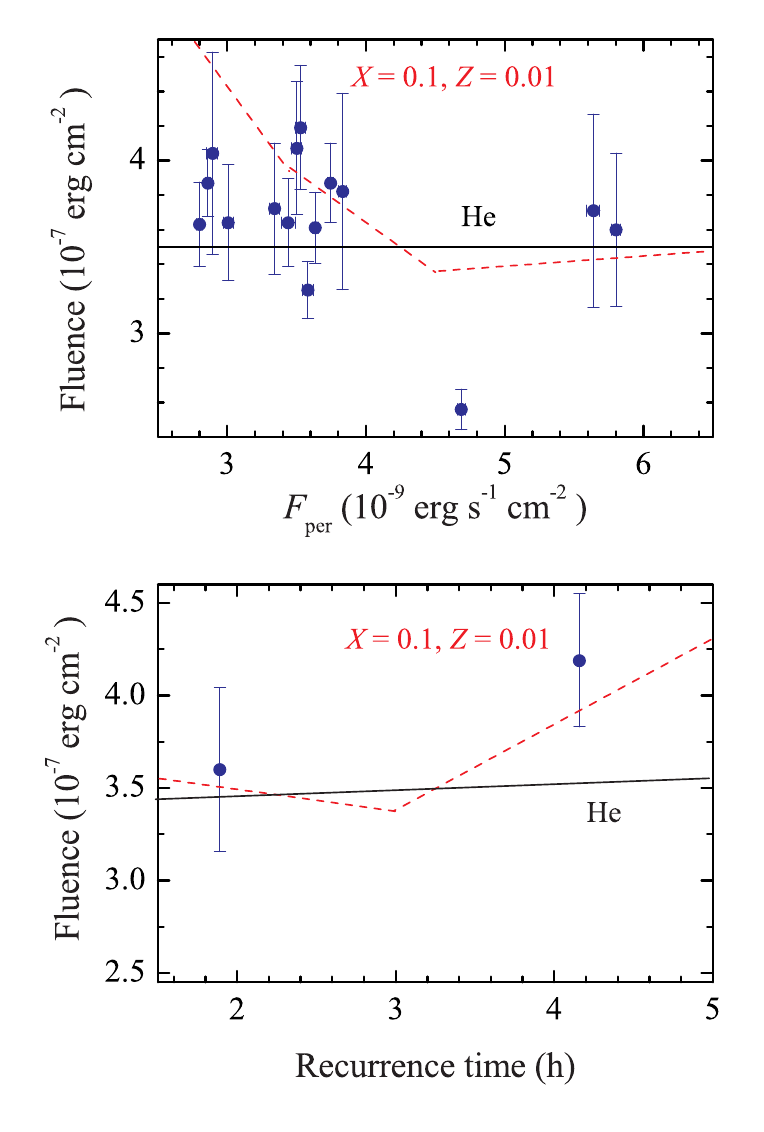}
\caption{\label{fig:recur} 
Dependences of the burst fluence on the persistent flux before the burst (top panel) and the recurrent time between bursts (bottom panel).  
The model curves for pure helium burning
(solid curve) as well as burning of helium with some mixture of hydrogen and heavy elements (dashed red curves) were computed by \citet{C03}. 
} 
\end{center} 
\end{figure}

The properties of the persistent state prior to the onset of all 16 short observed bursts from 
4U\,1820$-$30 are presented in Table\,\ref{tab2}.
\citet{Kajava.etal:14} demonstrated that X-ray bursts suitable for  constraining NS parameters  
have significant evolution of the normalization after the photospheric touchdown point. 
It approximately doubles from the touchdown to the moment when the flux is half the touchdown flux $F_{\rm td}$. 
In the considered bursts, the normalization reaches maximum values $K_{\rm max}$ at the flux of about
 $0.7\,F_{\rm td}$  and we used the ratio $K_{\rm max}/K_{\rm td}$ as a parameter describing the evolution of the normalization. 
This value is also shown for every burst in Table\,\ref{tab2}.    

  \begin{table*}
\centering   
\caption{Parameters of the persistent emission before the bursts  and burst fluences. 
\label{tab2} 
}
\begin{tabular}[c]{l c c c c c c c c c }
\hline
N & Obs. id. &  Date & 	$K_{\rm max}/K_{\rm td}$  &  HC  & SC & $\Gamma$ & $kT$    & $F_{\rm per}$ & Fluence  \\ 
 &  &  MJD &  &    & &  & (keV)   & ($10^{-9}$\,erg\,s$^{-1}$cm$^{-2}$) &  ($10^{-7}$\,erg\,cm$^{-2}$)\\
\hline 
1*  & 20075-01-05-00 & 50570.731795 & 1.53$\pm$0.15 & 0.66 & 1.15 & 2.33$\pm$0.04 & 2.33$\pm$0.04 & 3.58$\pm$0.04 
&3.25$\pm$0.17\\
2*  & 40017-01-24-00 & 52794.738826 & 1.63$\pm$0.13 & 0.92 & 1.22 & 1.93$\pm$0.03 & 2.18$\pm$0.08 & 2.80$\pm$0.02
 &3.63$\pm$0.25\\
3*  & 70030-03-04-01 & 52802.076265 & 1.55$\pm$0.14 & 0.93 & 1.24 & 1.83$\pm$0.03 & 2.01$\pm$0.06 & 2.86$\pm$0.02 
&3.87$\pm$0.19\\
4*  & 70030-03-05-01 & 52805.896358 & 1.73$\pm$0.14 & 0.84 & 1.25 & 2.06$\pm$0.03 & 2.41$\pm$0.06 & 3.74$\pm$0.02 
&3.87$\pm$0.23\\
5*  & 90027-01-03-05 & 53277.439257 & 1.42$\pm$0.12 & 0.61 & 1.13 & 2.45$\pm$0.05 & 2.31$\pm$0.03 & 3.64$\pm$0.02 
&3.61$\pm$0.21\\
6   & 94090-01-01-02 & 54948.821939 & 1.50$\pm$0.25 & 0.80 & 1.21 & 2.00$\pm$0.06 & 2.18$\pm$0.10 & 3.34$\pm$0.04
&3.72$\pm$0.38\\
7*  & 94090-01-01-05 & 54950.703513 & 1.38$\pm$0.14 & 0.94 & 1.26 & 1.86$\pm$0.03 & 2.19$\pm$0.10 & 3.44$\pm$0.05 
&3.64$\pm$0.26\\
8*  & 94090-01-02-02 & 54958.740672 & 1.73$\pm$0.18 & 1.07 & 1.27 & 1.69$\pm$0.05 & 1.93$\pm$0.16 & 3.01$\pm$0.04 
&3.64$\pm$0.34\\
9   & 94090-01-02-03 & 54956.775426 & 1.90$\pm$0.32 & 1.12 & 1.29 & 1.77$\pm$0.04 & 2.14$\pm$0.14 & 3.14$\pm$0.04 
&3.77$\pm$0.34\\
10  & 94090-01-04-00 & 54978.322182 & 1.83$\pm$0.34 & 1.14 & 1.28 & 1.70$\pm$0.05 & 2.00$\pm$0.17 & 3.50$\pm$0.04 
&4.07$\pm$0.39\\
11  & 94090-01-04-01 & 54978.495588 & 1.48$\pm$0.29 & 1.29 & 1.21 & 1.68$\pm$0.04 & 2.00$\pm$0.20 & 3.53$\pm$0.04 
&4.19$\pm$0.36\\
12* & 94090-01-05-00 & 54981.187938 & 1.62$\pm$0.13 & 1.13 & 1.35 & 1.71$\pm$0.03 & 2.33$\pm$0.09 & 3.83$\pm$0.03 
&3.82$\pm$0.57\\
13  & 94090-02-01-00 & 54994.534879 & 1.38$\pm$0.25 & 0.87 & 1.29 & 2.04$\pm$0.05 & 2.53$\pm$0.09 & 5.64$\pm$0.05 
&3.71$\pm$0.56\\
14  & 94090-02-01-00 & 54994.613713 & 1.18$\pm$0.13 & 0.82 & 1.24 & 2.12$\pm$0.03 & 2.51$\pm$0.05 & 5.80$\pm$0.03 
&3.60$\pm$0.44\\
15  & 96090-01-01-00 & 55624.881378 & 1.84$\pm$0.23 & 0.91 & 1.22 & 1.95$\pm$0.06 & 2.34$\pm$0.17 & 2.89$\pm$0.04 
&4.04$\pm$0.59\\
16  & 96090-01-01-02 & 55626.774306 & 1.27$\pm$0.14 & 0.51 & 1.01 & 2.55$\pm$0.04 & 2.08$\pm$0.03 & 4.69$\pm$0.03 
&2.56$\pm$0.12\\
 \hline
\end{tabular}
\begin{flushleft}{ 
Note: The persistent emission prior to the burst is characterized by the power-law index $\Gamma$ and blackbody temperature $kT$, the hard (HC) and the soft (SC) colours and the flux.  The bursts themselves are characterised by the ratio of the maximum blackbody normalization to that at the touchdown. 
The eight bursts, which are taken for analysis, are marked by a star.
Errors are $1\sigma$.
}\end{flushleft} 
\end{table*}

The burst fluences also contain potentially important information. \citet{C03} showed that pure helium
 burning gives burst fluences independent on the persistent flux, whereas a relatively small amount of hydrogen 
  ($X=0.1$) together with heavy element abundances $Z=0.01$  in the burning matter makes bursts 
  more energetic at low persistent fluxes.
The observed dependence is shown in Fig.\,\ref{fig:recur}\,(top panel), and it is rather
 flat supporting pure helium burning in this source.
The burst \# 16 with a very unusually low fluence took place in a soft-end of the hard 
persistent state and a rather high persistent flux (see Table \ref{tab2} and Fig.\,\ref{fig:color}). 
Possibly, part of the helium was burnt in the persistent regime before this burst 
 
\citet{C03} also predicted another dependence, between  the fluence and the recurrent time between bursts. 
Unfortunately, the {\it RXTE} observational sample consists of only two pairs of successive bursts.
Therefore, we can find the recurrent time for two bursts only, namely bursts \# 11 and 14.
The recurrence times are 4.16 h and 1.89 h, corresponding to the persistent fluxes 
of 3.5 and 5.8 $\times 10^{-9}$\,erg\,s$^{-1}$cm$^{-2}$.  
The shortening of the recurrence time is consistent with the results presented by \citet{Clark.etal:77} and \citet{Haberl.etal:87}. 
Formally, the result is in accordance to the hydrogen-helium mix with solar chemical 
abundances of heavy elements (see bottom panel of Fig.\,\ref{fig:recur}). 
However, it is obvious that it is just the result of small  numbers of the observed points (compare to the top panel of the same figure). 
The corresponding $\alpha$ values for these two bursts, 125$\pm$10 and 110$\pm$10 are in 
a good agreement with results presented by \citet{Haberl.etal:87}. 

\section{Application of the  direct cooling tail method}
\label{s:ctm}

The direct cooling tail method is based on an extended grid of the hot neutron star model atmospheres computed for a wide range of surface gravities ($\log g$ from 13.7 till 14.9 with the step 0.15),  relative luminosities (from the actual Eddington limit to 0.1 $L_{\rm Edd}$) and various atmosphere chemical compositions \citep{SPW12, Sul.etal:17}.
The emergent model spectra $\mathcal{F}_{\rm E}$ were fitted with a diluted blackbody $B_{\rm E}$ in the observed energy band 3$-$20 keV 
\be \label{mfit}
         \mathcal{F}_{\rm E} \approx w\, \upi B_{\rm E}(\fc T_{\rm eff}),       
\ee
where the model effective temperature $T_{\rm eff}$ is determined by the relative luminosity of the model $\ell = L/L_{\rm Edd}$, $T_{\rm eff} = \ell^{1/4}\,T_{\rm Edd}$.
Here $\sigma_{\rm SB}\,T_{\rm Edd}^4 = gc/\kappa_{\rm T}$ is a critical (Eddington) temperature on the NS surface, and $\kappa_{\rm T}\approx 0.2(1+X)$\,cm$^2$\,g$^{-1}$ is  the electron scattering opacity. 
The pairs $w,\fc$ were obtained for every computed model emergent spectrum. This kind of approximation was made because observed spectra of X-ray bursts are well fitted by blackbody spectra with temperature $T_{\rm BB}$ and normalization $K= R_{\rm BB}^2/d^2$ as parameters \citep[see e.g.][]{gallow08}. 
Here $d$ is the distance to the source. 
These observed spectrum fit parameters can be  easily compared with the model parameters $w$ and $\fc$ to obtain information about  physical NS radius $R$.

It is possible to derive the required  relation \citep{Sul.etal:17}
\be
     K = w\, \left(\frac{R(1+z)}{d}\right)^2 = w\,\Omega,
\ee
and to show that the observed dependence $K - F_{\rm BB}$ has to be fitted with the model
 dependence $w - w\fc^4\,\ell$ if the model of a passively cooling NS is acceptable for a given X-ray burst. 
The factor $w\fc^4$ is an inverted bolometric correction arising due to fitting the model spectra 
with equation (\ref{mfit}) \citep{Sul.etal:17}.
There are two fitting parameters for this procedure: $\Omega$, which is the angular dilution factor 
proportional to the solid angle occupied by the NS on the sky, and the observed Eddington flux 
$F_{\rm Edd,\infty}=L_{\rm Edd}/4\upi d^2 (1+z)^2$, which are combined to the observed Eddington temperature 
\be \label{eq:kTEdd}
kT_{\rm Edd,\infty} = kT_{\rm Edd}/(1+z) = 9.81\,(F_{\rm Edd,-7}/\Omega)^{-1/4} \,\mbox{keV}. 
\ee
This temperature is independent of the source distance $d$ and any errors in the absolute observed flux. 
Here $F_{\rm Edd,-7}$ is the observed Eddington flux $F_{\rm Edd,\infty}$ in units 
$10^{-7}$\,erg\,s$^{-1}$\,cm$^{-2}$ and $\Omega$ in units (km/10\,kpc)$^2$.  

Formally, only the crossing point of the constant $T_{\rm Edd,\infty}$ curve with the constant $\log g$ curve,
 for which the model dependence $w -w\fc^4\,\ell$ was computed, gives a correct solution. 
This is the reason, why the direct cooling tail method was introduced to cover all the $M-R$ plane. 
For every $M$ and $R$ pair the gravitational redshift $z$, the surface gravity $g$, and the Eddington luminosity
 $L_{\rm Edd}$ are known, if the chemical composition of the NS atmosphere is given. 
The model curve $w - w\fc^4\,\ell$ for a specific surface gravity $g$ could be interpolated among the nine computed
 $\log g$ dependences and used for fitting the observed $K - F_{\rm BB}$ curve. 
In the considered case the fitting parameters $\Omega$ and $F_{\rm Edd, -7}$ depend on the source distance 
$d$ only, which is, therefore, the only fitting parameter of the method.
The value of $\chi^2$ is used as a measure of the fitting accuracy.
Using this approach we can get a $\chi^2$ map in the $M-R$ plane and find the regions with the minimum of 
$\chi^2$ and the confidence regions containing the correct $(M,R)$ pair with the likelihood, say, of 68, 95 and 99 per cent. 

\begin{figure} 
\begin{center}
\includegraphics[width= 0.9\columnwidth]{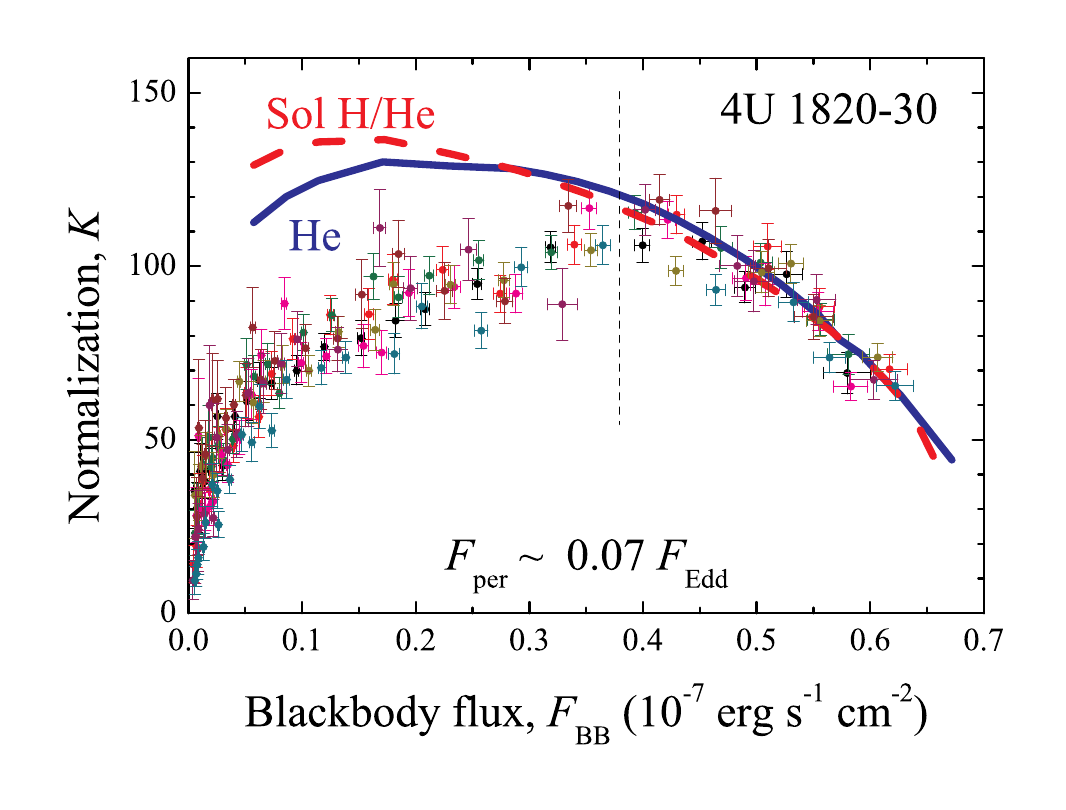}
\caption{\label{fig:KFbb} 
The observed $K-F_{\rm BB}$
dependence (circles with error bars) for
the cooling tail (after touchdown) of the selected bursts of 4U\,1820$-$30. The blue solid curve is the 
best-fitting theoretical dependence $w - w\fc^4\,\ell$ for pure helium and $\log g = 14.3$ 
corresponding to the selected data (points to the right of
the vertical dashed line with $F_{\rm BB} > 0.38\times 10^{-7}$\,erg\,s$^{-1}$\,cm$^{-2}$).
The fit parameters of the curve are $\Omega= 500$\,(km\,/\,10\,kpc)$^2$ and
$F_{\rm Edd} = 0.596\times 10^{-7}$\,erg\,s$^{-1}$\,cm$^{-2}$.
The red dashed curve is the best-fitting theoretical dependence for the solar hydrogen 
to helium abundance ratio and $\log g = 14.3$, 
with parameters $\Omega= 618$\,(km\,/\,10\,kpc)$^2$ and $F_{\rm Edd} = 0.596\times 10^{-7}$\,erg\,s$^{-1}$\,cm$^{-2}$.
} 
\end{center} 
\end{figure}

Selecting the part of the observed $K-F_{\rm BB}$ dependence, which is used for the fitting, could 
be crucially important for the method. 
The physically motivated truncation of the whole observed dependence at high fluxes is the touchdown point.
It is commonly accepted that  the photospheric radius equals to the NS radius at later burst cooling phases.
Therefore, we use the points with bolometric fluxes  $F_{\rm BB}$ less than the touchdown point flux.   
Determination of the cutoff boundary at low fluxes is much less  obvious. 
We expect that accretion can restart some time after the PRE phase and disturb the NS atmosphere, 
braking  the assumption of a passively cooling NS. 
The accretion rate during the burst actually can be even higher than before it \citep{Worpel.etal:13, Worpel.etal:15, Ji.etal:14}.
In our previous work \citep{nattila16} we tried to find the bursts, which took place at very low persistent luminosities
 to minimize the influence of accretion.  
This approach is valid if we want to find the most accurate limits on NS masses and radii. 
Here, however, our aim is different. 
We wish to demonstrate that the cooling tail method gives the correct chemical composition of the NS atmosphere.  
Because of the rather high persistent flux of the source (about 0.07\,$L_{\rm Edd}$) in the hard state, we have to
 limit our data to a  short part of the cooling phase at high flux levels where theoretical atmosphere models describe the data. 
At lower fluxes, deviations become obvious because of a likely significant influence of the  restarting accretion
 onto the observed burst properties (see Fig.~\ref{fig:KFbb}).

 \begin{figure} 
\begin{center}
\includegraphics[width= 0.9\columnwidth]{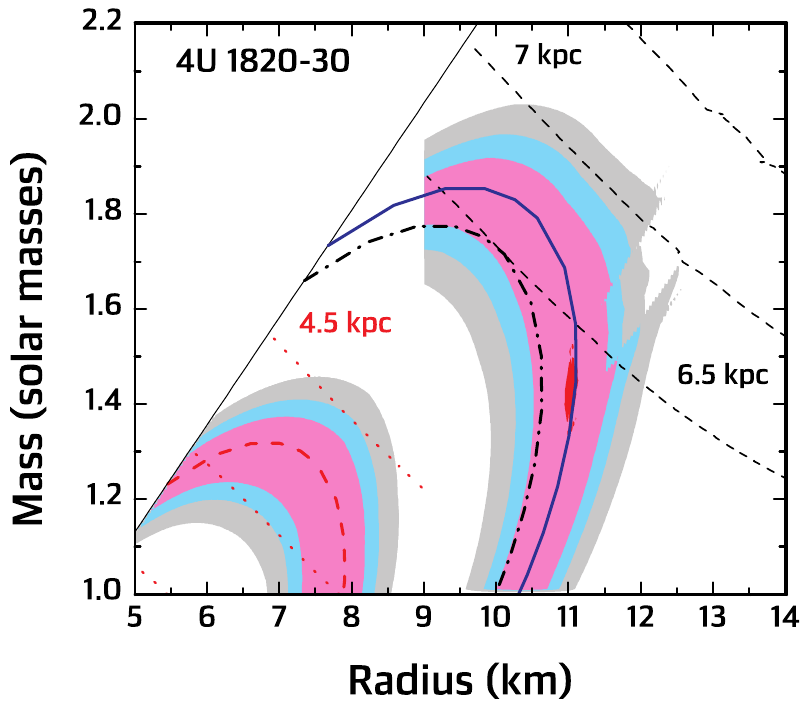}
\caption{\label{fig:chi2} 
The $\chi^2$ confidence regions (68, 90 and 99\% probabilities) in the mass-radius plane for
4U\,1820$-$30 obtained using the  direct cooling tail method for pure helium (right regions with larger radii)
and for the solar H/He abundance ratio  (left regions) model atmospheres.  
The best-fitting region for the He solution is marked by red. The minimum value of $\chi^2$ is  34.6 for 31 dof. 
The corresponding values for the best-fitting solution using the solar H/He ratio  are  36.9/31.
The blue solid curve  corresponds to the best-fitting $T_{\rm Edd, \infty}$ (equation \ref{eq:kTEdd}) for pure  helium, 
 the red dashed curve for the solar H/He mix atmospheres, and the black dot-dashed curve for the atmospheres computed 
 for $X=0.1, Y=0.9$ mix. 
All the curves were obtained for $\log g$=14.3. 
The black dashed curves correspond to the constant distances of 6.5, 7.0, and 7.5 kpc (solutions for pure helium), and 
the red dotted curves correspond to the constant distances of 3.5, 4.0, and 4.5 kpc (solution for solar H/He mix).
Solutions in the upper-left corner are forbidden because of general relativity effects.
}
\end{center} 
\end{figure}

We chose eight bursts for analysis, marked by stars in Table\,\ref{tab2}.  
They were chosen among others because they have the largest ratio  $K_{\rm max}/K_{\rm td}$ with sufficient 
quality of the fitting, making them likely to follow theoretical atmosphere models \citep{Kajava.etal:14,Poutanen.etal:14}.
Their observed dependences   $K - F_{\rm BB}$ are shown in Fig.\,\ref{fig:KFbb} together with the model curves 
$w - w\fc^4\,\ell$ computed for $\log g = 14.3$ and two chemical compositions,  pure helium and solar H/He mix. 
The found best-fitting parameters $\Omega$ and $F_{\rm Edd, -7}$ are also given.
The observed points at the fluxes below $4 \times 10^{-8}$\,erg\,s$^{-1}$\,cm$^{-2}$ deviate from 
the model curve significantly, and we excluded them from the analysis.
The used points occupy a very short (about 1~s) time interval during the bursts and this interval is shown in Fig.\,\ref{fig:cooling}.
We cannot expect very high accuracy from these data, but it could be enough to demonstrate 
that the pure helium models give an acceptable result. 
We note, that usage of all bursts give almost the same confidence regions at the $M - R$ plane, but with a more irregular structure.

The results obtained by  the  direct cooling method are shown in Fig.\,\ref{fig:chi2}. 
The confidence regions are situated along the whole $T_{\rm Edd, \infty}$ curve, found for the fixed $\log g = 14.3$, with  the 
$\chi^2$ minimum at  $R\approx 11$\,km and $M \approx 1.45$\,M$_\odot$ for pure He composition. 
The curves of constant distances  of 6.5, 7 and 7.5\,kpc are also shown. 
The most probable distance to the source is about 6.5\,kpc. 
This value is smaller than the commonly accepted distance to NGC\,6624 of 7.6$\pm$0.4\,kpc \citep{Kuulkers.etal:03}, 
although earlier investigations \citep{Vacca86} gave smaller distances, down to  6\,kpc.

The NS radius obtained here for a 1.5\,M$_\odot$ NS  is smaller than the radii obtained in our previous
 work \citep{nattila16} for other sources. 
We suggest two reasons for this difference. 
Firstly, if  the most distant acceptable solutions have a higher probability, our preferable solution will shift 
to larger radii $11 - 12$\,km and masses $1.8 - 2$\,M$_{\odot}$.  
Such radii are in a good agreement with those derived by \citet{nattila16}.  
The NS mass could be high enough because of the long accretion history. 
Secondly, the analysed bursts happened at a rather large persistent luminosity of about $0.07 L_{\rm Edd}$
\citep{Kajava.etal:14}, whereas \citet{nattila16} considered bursts at lower persistent luminosities $<0.02 L_{\rm Edd}$. 
Accretion during the bursts disturbs the NS atmosphere affecting the observed $K - F_{\rm BB}$ dependence not only
 at low fluxes, but even close to the touch-down. 
Our preliminary study (V.F. Suleimanov et al., in preparation) indicates that additional heating of the atmosphere
 top layers increases the colour correction factor $f_{\rm c}$ resulting in a shift of the $M-R$ solution to larger radii by about 1 km.

Many authors tried  to limit  mass and radius of the NS in 4U\,1820$-$30.  
\citet{vPL87} obtained an allowed region in the $M - R$ plane close to ours 
 when assuming a helium NS atmosphere and an anisotropy factor of $\xi = 1$. 
\citet{ST04} investigated the decay phase of 4U\,1820$-$30 bursts using their analytical
 representation of the $\fc$ and obtained results, which are close to ours: 
 $M \approx 1.3 - 2.0 M_\odot$, and $R\approx 11-13$\,km assuming the distance to the source between 5.8 and 7.0 kpc. 
The limitations derived by \citet{Kusm:11} are very wide ($M = 1.3 \pm 0.6$\,M$_\odot$, $R = 11^{+3}_{-2}$\,km),
 but they include our confidence regions, too. 
\citet{Guver.etal:10} reported very high accuracy of the NS mass and radius assuming pure He 
composition ($M = 1.58 \pm 0.06$\,M$_\odot$, $R = 9.1\pm0.4$\,km). 
However, their method was not self-consistent as shown by \citet{SLB10} and \citet{SPRW11}.
In particular, the variations of the normalization $K$ were ignored and only an averaged value was used. 
On the other hand, they took $\fc$ values from the model atmosphere computations ignoring the fact that $\fc$ and, therefore, $K$ as well have to vary with the decreasing burst flux. 
\citet{SLB10} remarked also that the solutions obtained by \citet{Guver.etal:10} were possible only with a probability of about 10$^{-7}-10^{-8}$.
The recent re-analysis by \citet{Ozel16} included additional correction factors and significantly higher 
uncertainties of the $K$ and $F_{\rm Edd, \infty}$ measurements together with much wider possible  errors for the distance. 
However, their new results are still plagued by the aforementioned inconsistencies because
 they kept the same approach as in \citet{Guver.etal:10}.
As a result of the (artifially) expanded measurement errors, they obtained much wider limits for the NS mass and radius 
(1.3$-$2.2 M$_\odot$ and 8$-$14\,km), which include our confidence regions as well.

Assuming solar H/He mix  for the NS atmosphere composition, we obtain NS radii below 9\,km
 (see lower left contours in Fig.\,\ref{fig:chi2}). 
Such radii are physically unrealistic  \citep[e.g., fig. 3 in][]{LP:16}. 
This solution also corresponds to a too small distance to the source of about 4\,kpc. 
Therefore, the data here are very strongly disfavouring the solar composition. 
This result is not affected  by the uncertainties in the colour correction caused by ongoing accretion, 
as it is more robust under small changes in the observables.  
This is because $F_{\rm Edd} \propto 1/(1+X)$\,  and as such, is a strong function of the hydrogen mass fraction.

Actually all the solutions between the solar H/He mix solution and the pure He solution
 shown in Fig.\,\ref{fig:chi2} are possible, if we consider the H/He ratio as a free parameter. 
For instance, we computed a set of model atmospheres with $X=0.1$ and $Y=0.9$ for a fixed $\log g = 14.3$ because $X=0.1$
is an upper limit for the hydrogen mass fraction in 4U\,1820$-$30 \citep{C03}. 
The model curve $w -w\fc^4\,\ell$ fits well the observed $K-F_{\rm BB}$ curve with the parameters 
$\Omega = 530$\,(km\,/\,10\,kpc)$^2$ and $F_{\rm Edd} = 0.6\times 10^{-7}$\,erg\,s$^{-1}$\,cm$^{-2}$.
The curve corresponding to the obtained $T_{\rm eff,\infty}$ is shown in Fig.\,\ref{fig:chi2}. We conclude that the existing
uncertainty of the chemical composition of the accreting matter in 4U\,1820$-$30 can shift the obtained limitations
on the NS radius to lower values by the value of the statistical error.

We would like to emphasize that the accurate NS mass and radius determination was not the main goal of this work: 4U\,1820$-$30 is not an ideal source for this purpose because of the high persistent flux level during the bursts. 
The main aim of the work was the demonstration of the potential power of the direct cooling method, which can give acceptable results
even for such a complex X-ray bursting NS. 
The method also confirmed that the accreting matter in 4U\,1820$-$30 is likely composed of pure He only.

\section{Summary}
\label{sec:summary}
 
We presented our re-analysis of {\it RXTE} observations of the ultracompact  X-ray bursting LMXB 4U\,1820$-$30 and 
confirmed that almost all detected X-ray bursts took  place during hard persistent states. 
Therefore, the cooling tail method can be used for analysis of these X-ray bursts to constrain the NS basic parameters. 
This source is not ideal for the method  because of a relatively high persistent accretion rate ($\sim 0.04-0.09 \,L_{\rm Edd}$) 
and a possible effect of the re-starting accretion on the spectral burst evolutions at the late cooling phases. 
We undertook this attempt mainly for demonstrating the method advantages  and confirming that it gives
 the  correct chemical composition of the accreting matter. 
The source is one of the best established binaries with a helium secondary and we demonstrated that for  pure helium composition 
the cooling tail method constrains the NS radius to be 11$\pm$1\,km for a low-mass ($M< 1.7$\,M$_{\odot}$)
 star and radii in the range 8$-$12\,km are possible for higher assumed masses.
We also find that the obtained solutions for the solar H/He mix result in a too small ($<$\,9\,km) 
NS radius and therefore can be excluded with a high degree of certainty. 
We have demonstrated using  the cooling tail method  that only a helium-dominated chemical composition  
of the NS atmosphere  in 4U\,1820$-$30  gives a NS mass and radius close to the canonical values for NSs.  
The uncertainties of the obtained $M$ and $R$ are, however,  too large to produce serious limitations on 
the equation of state in NS cores.  
The likely reason for that is the rather high persistent mass accretion rate, which can disturb the NS atmosphere and affect our
results.

\section*{Acknowledgments} 
 We acknowledge the anonymous referee, whose comments allowed  to improve the paper.
The work was supported by the German Research Foundation (DFG) grant WE 1312/48-1, 
the Magnus Ehrnrooth Foundation. The work was also supported by 
the Academy of Finland grants 268740 and 295114 (JP, JJEK),  
the University of Turku Graduate School in Physical and Chemical Sciences (JN), 
the Foundations' Professor Pool and the Finnish Cultural Foundation (JP), and Russian Science Foundation 
grant 14-12-01287 (VFS, SM, AL). We also thank to the International Space Science Institute (Bern, Switzerland) and COST
Action MP1304 NewCompStar for the organization of the group meeting and support of the
travel costs.       



\bibliographystyle{mnras}
\bibliography{xrb1} 

\bsp	
\label{lastpage}
\end{document}